\documentclass[prd,twocolumn,showpacs,preprintnumbers,amsmath,amssymb,superscriptaddress]{revtex4}
\usepackage{graphicx, epsfig, rotating, dcolumn, bm, xspace, color}

\def\lsim{\raisebox{-.4ex}{\rlap{$\sim$}} \raisebox{.4ex}{$<$}}

\begin{document}

\preprint{TIFR/TH/13-05}

\title{Testing Times for Supersymmetry: Looking Under the Lamp Post}

\author{Amol Dighe}
\affiliation{Department of Theoretical Physics, Tata Institute of 
Fundamental Research, Mumbai 400~005, India.}
\author{Diptimoy Ghosh}
\affiliation{INFN, Sezione di Roma, Piazzale A. Moro 2, I-00185 Roma, 
Italy.}
\author{Ketan M. Patel}
\affiliation{Department of Theoretical Physics, Tata Institute of 
Fundamental Research, Mumbai 400~005, India.}
\author{Sreerup Raychaudhuri}
\affiliation{Department of Theoretical Physics, Tata Institute of 
Fundamental Research, Mumbai 400~005, India.}

\begin{abstract} 
\noindent We make a critical study of two highly-constrained models of 
supersymmetry --- the constrained minimal supersymmetric standard model 
(cMSSM), and the non-universal Higgs mass model (NUHM) --- in the light 
of the 125-126 GeV Higgs boson, the first observation of $B_s \to 
\mu\mu$ at the LHCb, and the updated $B \to \tau \nu$ branching ratio at 
BELLE. It turns out that these models are still allowed by the 
experimental data, even if we demand that there be a light stop with 
mass less than 1.5~TeV. The only significant effects of all these 
constraints are to push the mass of the light stop above $\sim 500$ GeV, 
and to prefer the universal trilinear coupling $A_0$ to be large and 
negative. We calculate the Higgs boson branching ratios to $WW, ZZ, 
\tau\tau$ and $\gamma\gamma$ in these models and show that improved 
experimental limits on these could put them to the most stringent 
experimental tests yet.
\end{abstract} 
\maketitle

The recent discovery~\cite{Higgs_search}, at the CERN LHC, of a boson 
with mass around 125-126~GeV and properties closely resembling those of 
the Standard Model (SM) Higgs boson~\cite{EWSB, GSW}, matches well with 
a common prediction of all supersymmetric models~\cite{SUSY}, viz., the 
existence of a light Higgs boson with mass in the $100 - 135$~GeV range. 
However, as this could be sheer coincidence, only the discovery of a 
superpartner of one of the known SM particles would really establish the 
existence of supersymmetry (SUSY). Now, the elevation of the light Higgs 
boson mass to a value above $M_Z \approx 91$~GeV in supersymmetric 
models is known~\cite{SUSY} to be primarily due to radiative corrections 
involving the top quark ($t$) and the light and heavy stop states 
($\tilde{t}_1, \tilde{t}_2$). It follows, therefore, that the discovery 
of a light Higgs boson candidate in the upper part of the SUSY-allowed 
range may be considered as a hint of the existence of a rather low-lying 
stop state ($\tilde{t}_1$) --- perhaps light enough to be discovered at 
the LHC. We cannot, however, be too sure about this, for SUSY models can 
be practically indistinguishable from the SM if we take the so-called 
{\it decoupling limit}, where all the superpartners are too heavy to be 
detected at the LHC (or any terrestrial experiment conceivable in the 
near future).

Faced with this situation, we approach the problem in a pragmatic way, 
viz. we focus on that part of the large SUSY parameter space that could 
support a light stop state {\it discoverable at the LHC}. Though this 
does resemble the proverbial search under the lamp post, at the present 
juncture, it seems the most reasonable thing to do. The question then 
arises as to how light the $\tilde{t}_1$ state must be. An exact answer 
is, of course, a matter of detail, but one can form a crude estimate 
based on the fact that the production of $\tilde{t}_1 \bar{\tilde{t}}_1$ 
pairs at the LHC is dominated by strong interactions and hence depends 
essentially on the mass of the $\tilde{t}_1$ alone. The variation of the 
$\tilde{t}_1 \bar{\tilde{t}}_1$ pair-production cross-section with 
$m_{\tilde{t}_1}$ at the 14-TeV LHC is easily 
calculated~\cite{Prospino}, and this falls from a few pb at 
$m_{\tilde{t}_1} \simeq 300$~GeV to about 10~ab for $m_{\tilde{t}_1} 
\simeq 1.5$~TeV. As a typical estimate of the integrated luminosity of 
the LHC is about 3~ab$^{-1}$, the latter value corresponds to around 30 
$\tilde{t}_1 \bar{\tilde{t}}_1$ pairs --- which may be just enough for a 
discovery. Accordingly, we take $m_{\tilde{t}_1} \leq 1.5$~TeV as a 
reasonable criterion for a possible LHC discovery of the light stop.

This light stop criterion is not, however, the only restriction on the 
parameter space of the SUSY model in question. The requirement of a 
light Higgs state in the range $122-129$~GeV, coupled with the recent 
observation of the rare decay $B_s \to \mu^+\mu^-$~\cite{LHCb:2012ct}, 
the updated $B \to \tau \nu$ branching ratio at 
BELLE~\cite{Adachi:2012mm}, and the negative results of all direct 
searches for new particles at the LHC~\cite{PDG}, taken together, make a 
formidable combination, imposing tight restrictions on models of new 
physics.  Among the most severely affected are models where the 
electroweak sector differs from that of the SM. All SUSY models fall in 
this class. In an article~\cite{Ghosh:2012dh} written last summer, just 
before the announcement of the new boson discovery, two of the present 
authors had shown that even the most restrictive model of supersymmetry, 
viz. the constrained Minimal Supersymmetric SM (cMSSM)~\cite{SUSY}, was 
affected by the then-available data only marginally --- at the periphery 
of its parameter space --- leaving most of its parameter choices viable. 
The new data obtained since then have, in the span of a few months, 
changed the situation quite dramatically --- the allowed parameter space 
now appears to be severely restricted~\cite{newdata_implications}.

Since the new data are all compatible with the SM, they may be expected 
to drive SUSY towards the decoupling limit. This, of course, runs
contrary to the light stop scenario desired by us. In the current 
article, we examine whether the cMSSM can accommodate these conflicting 
requirements. Moreover, we also consider a popular extension of the 
cMSSM, known as the Non-Universal Higgs Mass (NUHM) 
model~\cite{Ellis:2002iu}, which could, in principle, evade some of the 
constraints which affect the cMSSM adversely. Both the models in 
question have been widely discussed in the 
literature~\cite{global-cmssm, global-cmssm-nuhm}, and do not require to 
be extensively reviewed in a focussed study such as the present one. Below, we 
just summarize the principal motivation(s) and the parameter choices.

The cMSSM is a theoretically well-motivated supersymmetric model with 
the least number of parameters, and consequently has the highest 
predictivity of all low-energy SUSY models. The free parameters of this 
model are (i) the universal fermion mass $m_{1/2}$ at the grand 
unification (GUT) scale, (ii) the universal trilinear coupling $A_0$ at 
the GUT scale, (iii) the ratio $\tan\beta$ of vacuum expectation values 
of the two Higgs doublets at the electroweak scale, and (iv) the 
universal scalar mass $m_0$ at the GUT scale. In addition, we can choose 
the sign of the Higgs mixing parameter $\mu$ to be positive or negative.

Predictivity is the most appealing feature of the cMSSM, though it also 
makes it more vulnerable to all sorts of experimental constraints. 
Therefore, even if the cMSSM is ruled out by the data --- or, more 
likely, pushed to its decoupling limit --- it would be logical to look 
for its close cousins, {\it i.e.} models which contain at least some of 
the more attractive features of the cMSSM. The NUHM model is perhaps the 
most natural of these choices, for all that it does is to relax one of 
the cMSSM conditions and make the two Higgs mass parameters $M_{h_1}$ 
and $M_{h_2}$ different from the universal scalar mass $m_0$ at the GUT 
scale.

In our numerical studies, we study the independent variation of these 
parameters in the ranges
\begin{eqnarray*}
m_{1/2} \in [0,~2]~{\rm TeV},& & m_{0} \in [0,~5]~{\rm TeV},  \\
A_0 \in [-10,~10]~{\rm TeV},& & \tan\beta \in [5,~60],  \\
M_{h_1} \in [0,~5]~{\rm TeV},& & M_{h_2} \in [0,~5]~{\rm TeV}, 
\end{eqnarray*}
which are compatible with theoretical considerations and cover the 
entire region which allows for a light stop $\tilde{t}_1$ of mass below 
1.5~TeV. We also restrict ourselves to $\mu > 0$, since only the 
positive sign of $\mu$ can ameliorate the $3.6\sigma$ disagreement of 
the SM with the measurement of the anomalous magnetic moment $(g-2)_\mu$ 
of the muon~\cite{PDG}. However, as it is still not possible to be 
compatible with the $(g-2)_\mu$ measurement to better than 95\% 
C.L.~\cite{gminus2}, we choose to be conservative and do not impose this 
particular constraint in any other way. Having chosen the parameter 
ranges, then, we make a random scan over the entire range of these 
variables. For each set of input parameters, we use the public domain 
software {\sc SUSYHIT}~\cite{Djouadi:2006bz} to compute the values of 
the entire supersymmetric spectrum and couplings at the respective 
scales at which the experimental data are available. These are then 
taken as inputs to calculate a slew of high- and low-energy observables 
for which experimental data are available. For this part, we use the 
software {\sc SuperISO}~\cite{Mahmoudi:2008tp} and {\sc 
FeynHiggs}~\cite{feynhiggs}.

\begin{center}
\begin{figure*}[htbp!]
\includegraphics[width=0.75\textwidth]{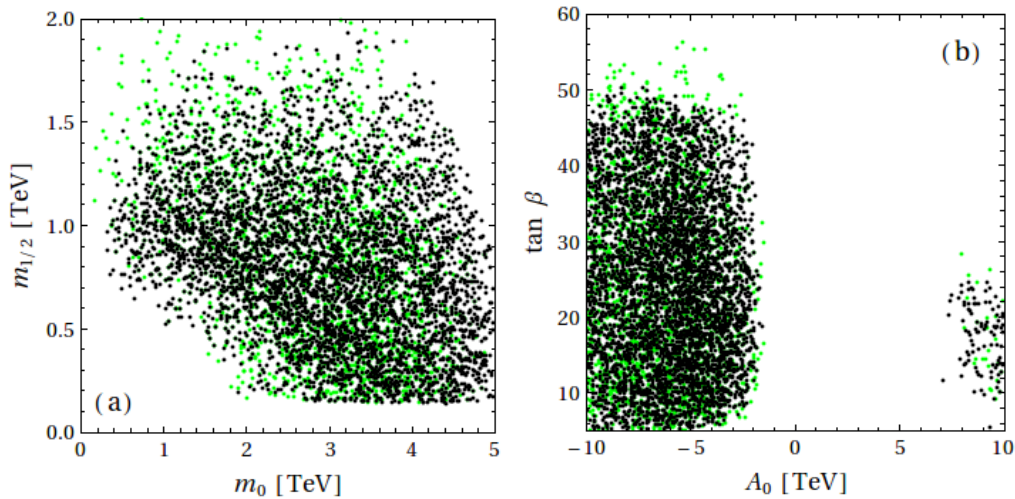}
\caption{Allowed regions in the parameter space for $\mu > 0$. We 
consider two sections of the parameter space, viz. (a) the 
$m_{1/2}$--$m_0$ plane, and the (b) $A_0$--$\tan\beta$ plane. The black 
dots indicate the cMSSM, while the green (grey) ones indicate the NUHM.}
\label{fig:parameters}
\end{figure*}
\end{center}

\vspace*{-0.35in}      
In order to constrain the parameter space, we first impose the 
``theory'' constraints, {\it i.e.} the obvious requirement that the 
electroweak symmetry be broken at precisely the electroweak scale, and 
the requirement that the lightest sparticle be neutral (which arises 
from its identification as a dark matter candidate). However, taking a 
conservative stance again, we do not impose constraints arising from the 
relic density of this neutralino, since such constraints can be evaded 
in scenarios with non-thermal dark matter~\cite{nonthermalDM}. In 
addition, we use the following constraints, many of which have been 
recently updated (mostly during 2012):

(i) The mass of the lightest Higgs boson $M_{h}$ lies in the narrow 
range $122.5~{\rm GeV} < M_{h} < 129.5~{\rm GeV}$. This range includes 
the 95\% C.L. errors from the ATLAS and CMS data~\cite{Higgs_search} as 
well as a 1.5~GeV theoretical uncertainty~\cite{Arbey:2012bp}.

(ii) The branching ratios ${\cal B}$ for some rare $B$ decays are 
constrained as follows, at 95 \% C.L. \cite{LHCb:2012ct,Asner:2010qj,Yook}
\begin{eqnarray*} 
1.1 \times 10^{-9} < &{\cal B}&\!\!(B_s \to \mu^+\mu^-) < 6.4 \times 
10^{-9}  \ , \\
&{\cal B}&\!\!(B_d \to \mu^+\mu^-) < 9.4 \times 10^{-10}  \ ,  \\
2.6\times 10^{-4} < &{\cal B}&\!\!(B \to X_s\gamma) \ \ \ < 4.5\times 
10^{-4} \ , \\
0.44\times 10^{-4} < &{\cal B}&\!\!(B\to \tau \nu_\tau) \ \ \ \ < 
1.48\times 10^{-4} \ .
\end{eqnarray*}       
As a matter of fact, the first measurement of ${\cal B}(B_s \to \mu 
\mu)$ has recently been reported at the LHCb~\cite{LHCb:2012ct}. Also, 
the measurement of ${\cal B}(B \to \tau \nu)$ with hadronic tag has 
recently been updated by BELLE~\cite{Adachi:2012mm} and brings the value 
of this ratio to a much smaller value than earlier measured. This not 
only relaxes the tension of this measurement with the SM prediction, but 
also dissipates the strong constraints it had put on the 
cMSSM~\cite{taunu-paper-island}. Here we take the BELLE average of 
${\cal B}(B \to \tau \nu)$~\cite{Yook}.

(iii) Another low energy measurement arises from the 2006 measurement, at
the DA$\Phi$NE facility at Frascati, of the process $K^+ \to \mu^+ \nu_\mu$.
Taking the 95\% C.L. limits, this gives us \cite{KLOE}
\begin{eqnarray*}
0.6331 < {\cal B}(K^+ \to \mu^+ \nu_\mu) < 0.6401 \ ,
\end{eqnarray*} 
which can be very restrictive for a certain class of electroweak models.                   
                          
(iv) We also impose the 95\% C.L. lower bounds on the masses of 
sparticles, as listed by the Particle Data Group (PDG)~\cite{PDG}. These 
are given below, with all masses being measured in GeV:
$$ 
\begin{array}{lll}
M_{\widetilde{\chi}_1^0} > 46 \; ,  & 
m_{\widetilde{q}} > 1100 \; ,    &
m_{\widetilde{\ell}} > 107 \; ,  \\
M_{\widetilde{\chi}_2^0} > 116 \; , &
m_{\tilde{t}_1} > 95.7 \; ,   &
m_{\widetilde{\nu_\ell}} > 94 \; ,  \\
M_{\widetilde{\chi}_1^\pm} > 94 \; , &
m_{\widetilde{b}_1} > 89 \; ,    &
m_{\widetilde{\tau}_1} > 81.9 \; , \\
M_{\widetilde{g}} > 500 \; . & & 
\end{array} $$
Here $\widetilde{q}$ and $\widetilde{\ell}, \widetilde{\nu}_\ell$ refer 
to superpartners of the first two generations. There are more specific 
constraints on the gluino mass $M_{\widetilde{g}}$, which depend on the 
mass $m_{\tilde{t}_1}$ of the light stop~\cite{PDG}; we have verified 
that these are satisfied for our allowed points.

Note that in selecting the allowed points in our random scan, we do not 
treat the early measurements of the Higgs boson branching ratios as 
constraints, since these are based on a rather small sample of Higgs 
boson decay events and, consequently, have large error bars. Instead, we 
shall presently argue that these measurements could constitute stringent 
tests for the models in question as more data become available.
                                                            
Taking all these constraints into account, the allowed points in our 
random scan are shown in Figure~\ref{fig:parameters}. The left panel --- 
marked (a) --- shows the $m_0$--$m_{1/2}$ plane, while the panel on the 
right --- marked (b) --- shows the $A_0$--$\tan\beta$ plane. We should 
note that all the parameters vary in these plots, {\it e.g.} in the 
panel (a), $A_0$ and $\tan\beta$ are not held fixed but are allowed to 
vary over the full ranges mentioned above. A glance at the panel~(a) 
shows that the two models, viz. cMSSM and NUHM, hardly differ so far as 
the allowed region is concerned. Only in the so-called stau 
co-annihilation region, where $m_0$ is small and $m_{1/2}$ is large, 
does the NUHM enjoy a slightly more viable status than the cMSSM. 
However --- and this is mostly due to the Higgs boson mass measurement 
--- both the models are severely constrained in the low-$m_0$ and 
low-$m_{1/2}$ region. Unfortunately, this is also the region in which 
large signals for superpartners are predicted at collider machines, such 
as the LHC, so this result is unfavourable to the school of thought that 
supersymmetry is ``around the corner''. The paucity of points in the 
``decoupling'' region of large-$m_0$ and large-$m_{1/2}$ is a direct 
consequence of imposing the restriction $m_{\tilde{t}_1} \leq 1.5$~TeV. 
However, the large number of allowed points in the central region of the 
panel clearly indicates that these two models --- with all their economy 
of parameter choices --- are still capable of explaining all the 
experimental data, {\it as well as} predicting a light stop. This 
conclusion is also reflected in the panel~(b), where it is clear that 
large negative values of $A_0$ are favored, but otherwise the only real 
restrictions are on large values of $\tan\beta$. These arise from the 
$\tan^6\beta$ sensitivity of ${\cal B}(B_s \to 
\mu^+\mu^-)$~\cite{Lunghi:2006uf}, coupled with the regime imposed by 
our artificial cutoff $m_{\tilde{t}_1} \leq 1.5$~TeV. In the cMSSM, this 
forces us to have $\tan\beta \lsim 45$, but the NUHM can still have 
allowed points with larger $\tan\beta$ values.

\begin{figure}[hp]
\vspace*{0.1in}
\begin{center}
\includegraphics[width=0.3\textwidth]{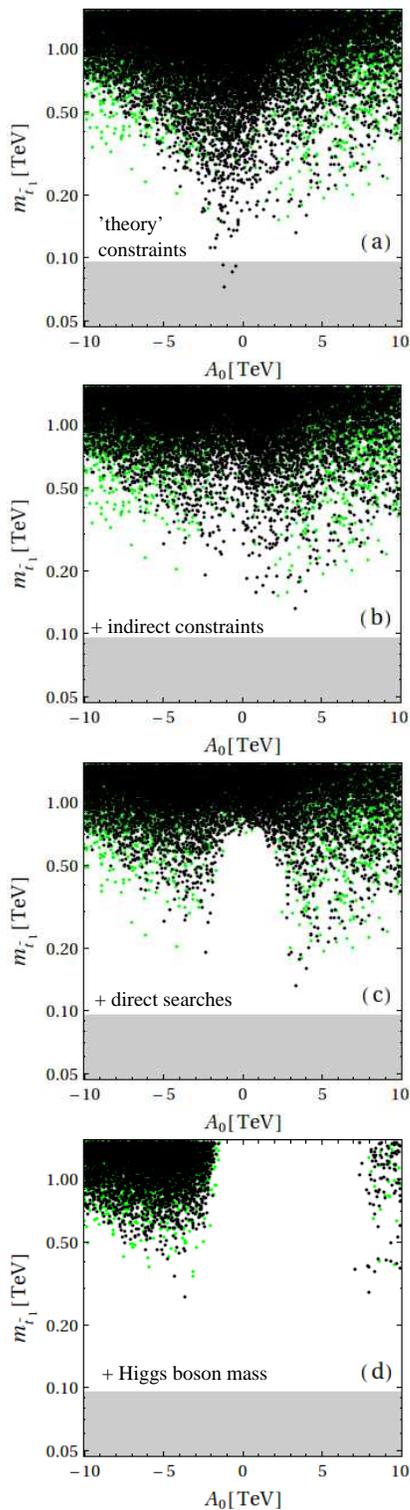} 
\caption{Illustrating the correlation between the mass of the light stop 
and the $A_0$ parameter, as different sets of constraints are imposed 
successively. Each panel includes the constraints from the previous 
panel. The shaded region represents the LEP constraint on 
$m_{\tilde{t}_1}$. Other conventions followed are exactly the same as in 
Fig.~\ref{fig:parameters}.  }
\label{fig:lightstop}
\end{center}
\end{figure}

It is interesting to ask why large negative values of $A_0$ are 
preferred. This is because the measured mass of the Higgs boson $M_h 
\simeq$~125-126~GeV is sufficiently removed from the decoupling limit 
--- typically quoted~\cite{SUSY} as $M_h \approx 119$~GeV for $A_0 = 0$ 
--- to require a light stop $\tilde{t}_1$ to explain that deviation. In 
order to sustain a light stop, while simultaneously keeping the squarks 
of the first two generations heavy, we require a large ``seesaw'' effect 
in the stop mixing matrix, and this happens when large negative values 
of $A_0$ appear in the off-diagonal terms of that matrix.

This is nicely illustrated in Fig.~\ref{fig:lightstop}, where we have 
shown the scatter of parameter choices with the mass $m_{\tilde{t}_1}$ 
plotted against the $A_0$ parameter, and the effect of different classes 
of constraints on the allowed region in this plot. In panel~(a), we 
impose only ``theory'' constraints, which are not only indifferent to 
the sign of $A_0$, but also allow a stop lighter than 100~GeV in the $A_0 
\approx 0$ region. In panel~(b), we impose, in addition, the indirect 
constraints arising from low-energy measurements, including B-decays, 
and this immediately affects the $A_0 \approx 0$ region, driving the 
stop mass above 120~GeV. Imposition of the direct search constraints, 
especially the requirement that $m_{\widetilde{q}} > 1.1$~TeV for the 
first two generations~\cite{PDG}, cleans up the $A_0 \approx 0$ region 
in panel~(c) much more efficiently, though a few points with a stop mass 
between 120-200~GeV still survive. Here, too, both signs of $A_0$ are 
equally preferred. However, when we impose the Higgs boson mass 
constraint, the allowed parameter space in panel~(d) is affected quite 
severely. Not only are small values of $A_0$ completely washed-out, with 
$A_0$ being driven below $-2$~TeV or above 8~TeV, but the stop mass is 
essentially pushed above 500~GeV, with only a few outlying points in the 
range 250-500~GeV (which may very well be due to specially fine-tuned 
cancellations). Thus there is no real surprise in the fact that 
experimental searches for stops lighter than 550~GeV at the 7-8~TeV LHC 
have come up empty-handed~\cite{ATLASstop}. On the other hand, it is 
encouraging that there is a profusion of allowed points with a 
$\tilde{t}_1$ mass in the range 600~GeV--1~TeV --- which, as we have 
argued, could easily turn out to be a happy hunting ground at the 14~TeV 
LHC.
 
We now come to the final test, viz, whether the models in question can 
predict the observed branching ratios of the Higgs boson as observed in 
the ATLAS~\cite{ATLAS-Moriond} and CMS~\cite{CMS-Moriond-WW,
CMS-Moriond-ZZ,CMS-Moriond-tau,CMS-Moriond-digamma} data. Of course, what 
is observed at the LHC is not the branching ratio per se, but a number 
of events of a certain kind which can be identified as arising from the 
decay of a Higgs boson resonance. This corresponds to the convolution of 
the integrated luminosity ${\cal L}$ with a production cross-section and 
a branching ratio, {\it i.e.} for a particle pair $X\bar{X}$, we have
\begin{equation}
N_{X\bar{X}}
= {\cal L} \times \sigma(pp \to h^0) \times {\cal B}(h^0 \to X\bar{X}) \; .
\end{equation}
We then define the so-called {\it signal strength} 
\begin{equation}
\mu_{X\bar{X}} \equiv \frac{N_{X\bar{X}}}{N_{X\bar{X}}^{\rm SM}} 
= \frac{\sigma(pp \to h^0)}{\sigma_{\rm SM}(pp \to h^0)} \times 
\frac{{\cal B}(h^0 \to X\bar{X})}{{\cal B}_{\rm SM}(h^0 \to X\bar{X})} \; ,
\end{equation}
where the label SM indicates the SM predictions with $m_h = 126$ GeV. 
The Higgs production cross-sections are calculated using {\sc FeynHiggs} 
and the branching ratios are calculated using the software {\sc HDecay}, 
which is incorporated in the {\sc SUSYHIT} package and is used in 
conjunction with the supersymmetry parameters allowed by the constraints 
described above. Our results can then be compared with the latest 
experimental values of $\mu_{X\bar{X}}$ given by the two experimental 
collaborations~\cite{ATLAS-Moriond,CMS-Moriond-WW,
CMS-Moriond-ZZ,CMS-Moriond-tau,CMS-Moriond-digamma}, which are listed in 
Table~\ref{tab:mews} below.
     
\begin{table}[!htbp]
\begin{center}
\begin{tabular}{lcccc}
\hline
    & \hspace*{0.2in} &  ATLAS          & \hspace*{0.2in} &  CMS     \\ 
\hline \\ [-2mm]
$\mu_{WW}$          & &  $1.0 \pm 0.3$ \cite{ATLAS-Moriond}  & & 
$0.76 \pm 0.21$ \cite{CMS-Moriond-WW} \\ [1mm]
$\mu_{ZZ}$          & &  $1.5 \pm 0.4$ \cite{ATLAS-Moriond} & & $0.91_{-0.24}^{+0.30}$ ~~\cite{CMS-Moriond-ZZ} \\ [1mm]
$\mu_{\tau\tau}$    & &  $0.8 \pm 0.7$ \cite{ATLAS-Moriond} & & 
$1.1 \pm 0.4$ ~~\cite{CMS-Moriond-tau} \\ [1mm]
$\mu_{\gamma\gamma}$& &  $1.6 \pm 0.3$ \cite{ATLAS-Moriond} & & $0.78_{-0.26}^{+0.28}$ ~~\cite{CMS-Moriond-digamma} \\[1mm]
\hline
\end{tabular}
\caption{Experimental data on Higgs boson signal strengths, as reported 
in March 2013 by the ATLAS~\cite{ATLAS-Moriond} and CMS~\cite{CMS-Moriond-WW,
CMS-Moriond-ZZ,CMS-Moriond-tau,CMS-Moriond-digamma} 
Collaborations. For $\mu_{\gamma\gamma}$ we use the multivariate analysis
result  from CMS \cite{CMS-Moriond-digamma}.}
\label{tab:mews}
\end{center}
\vspace*{-0.3in} 
\end{table}
        
From Table~\ref{tab:mews}, it is clear that at 95\%~C.L. all the 
measurements are compatible with a SM Higgs boson. We also note that 
these results are based on almost all the data collected in the 7--8~TeV
run of the LHC, and hence, no substantial improvement may be expected
till the energy/liminosity upgrade is in place. 

\begin{center}
\begin{figure*}[ht]
\includegraphics[width=0.75\textwidth]{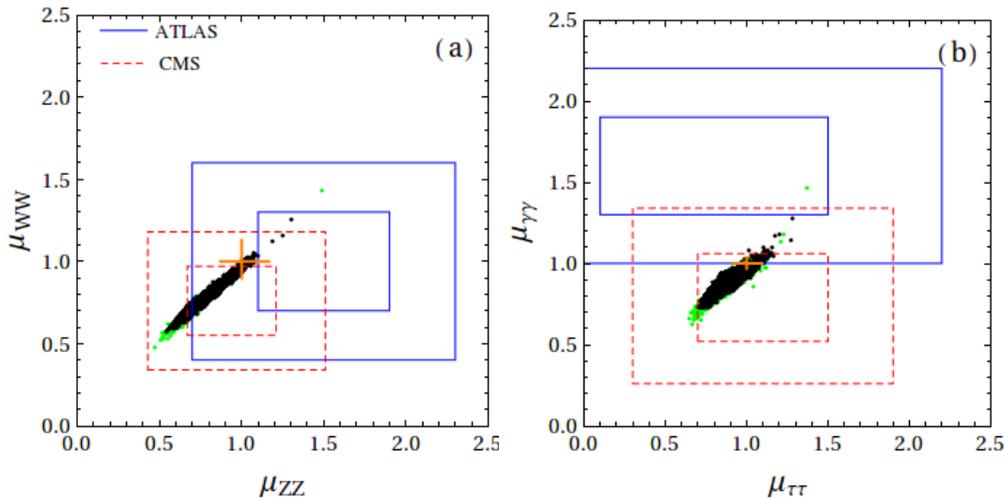}
\caption[]{Correlation plots for the signal strengths of the Higgs boson 
in different channels, vis-a-vis ATLAS and CMS bounds. SUSY conventions 
follow those of Fig~\ref{fig:parameters}. In each panel, the SM 
prediction lies at the crossing of the two short (orange) lines, which 
denote the theoretical uncertainty due to the spread in Higgs boson 
mass.}
\label{fig:decays}
\end{figure*}
\vspace*{-0.4in}
\end{center}

With these circumstances in mind, we now make a comparison between the 
signal strengths predicted in the cMSSM and the NUHM versus the 
experimental data. These results are shown in Fig.~\ref{fig:decays}. In 
the panel on the left --- marked~(a) --- we plot $\mu_{WW}$ versus 
$\mu_{ZZ}$, with black dots indicating the cMSSM and green (grey) dots 
indicating the NUHM. The blue (solid line) boxes indicate the ATLAS 
limits at $1\sigma$ and $2\sigma$ respectively, while the red (dashed 
line) boxes indicate the CMS limits, again at $1\sigma$ and $2\sigma$ 
respectively. The SM prediction lies at (1.0, 1.0), i.e. at the crossing 
of the two short (orange/grey) lines, whose lengths correspond to the 
uncertainties arising from the allowed range in Higgs boson mass.  It is 
immediately obvious that the CMS data are slightly more restrictive than the 
ATLAS data. If we take the bounds at $2\sigma$, then it is obvious that 
the full scatter of points in both the cMSSM and NUHM are allowed by the 
data, as is, of course, the SM. On the 
other hand, if we consider the $1\sigma$ constraints, then the SM is 
disfavoured, whereas the two supersymmetric models survive. 

The comparison changes a little if we turn to the panel on the right --- 
marked~(b). The conventions for this plot are exactly the same as 
before, the only difference being that we have plotted the more 
controversial $\mu_{\gamma\gamma}$ versus $\mu_{\tau\tau}$, the former 
being highly sensitive to new physics contributions. Here the ATLAS and 
CMS results are somewhat different. The SM point as well as the scatter 
of points in the two supersymmetric models lie in the heart of the CMS-allowed
region, whereas for the ATLAS result, the SM lies at the 2$\sigma$ border,
while most of the supersymmetric points lie further out. Thus, in the 
ATLAS data, at least, there is some hint of new physics, but the CMS
data appears to confound any such speculations. 

It is quite clear from this study of the signal strengths that --- as of 
now --- both the cMSSM and NUHM survive all the experimental tests, even 
if we demand a light stop state discoverable at the LHC. In fact, they 
do as well, if not better, than the SM --- as is only to be expected 
from models with more free parameters. The predictions of these models 
in their allowed ranges may deviate significantly from the SM 
predictions; however the smallness of the scatter compared to the 
experimental errors in both the plots in Fig.~\ref{fig:decays} shows 
that, the present status of these models is not so different from that 
of the SM. Indeed, in both the panels of Fig.~\ref{fig:decays} the SM 
point lies right in the middle of a dense clustering of allowed SUSY 
points. This indicates that these two models can mimic the SM 
predictions for these signal strengths {\it even with a light stop} of 
mass $m_{\tilde{t}_1} \leq 1.5$~TeV. It is thus much too early to write 
off any model of supersymmetry, even the highly restrictive cMSSM.

As regards the future, one can only speculate. Obviously, when the 
results from the analysis of more data become available, the boxes in 
Fig.~\ref{fig:decays} will shrink, corresponding to a reduction in the 
experimental errors. Such changes are often accompanied by changes in 
the central value(s) as well, which would appear on the plots as a 
translation of the box(es) in some direction. Three distinct 
possibilities may now be considered.

(i) The new boxes may zoom in to the SM prediction. In this case our 
plots in Fig.~\ref{fig:decays} show that there are always points in the 
parameter space of the cMSSM and NUHM which will give the same 
$\mu_{X\bar{X}}$ predictions as the SM. Thus, these models will survive 
--- and even predict a discoverable light stop --- though with even more 
severe constraints on the parameter space.

(ii) The shrinking boxes may move away from the SM predictions, but stay 
within the scatter of allowed SUSY points shown in 
Fig.~\ref{fig:decays}. This, would, of course, form a powerful argument 
for these minimal versions of SUSY and would certainly enhance our 
expectations for a light stop discovery.

(iii) In addition, we must also consider the possibility that the new 
data will disfavour {\it both} the SM and these two highly-predictive 
SUSY models --- in which case one would have to invoke less restrictive 
versions of SUSY (e.g. the 19-parameter 
pMSSM~\cite{newdata_implications}) or other forms of new physics to 
explain the results.

Before concluding, we briefly consider the case when the ``under the 
lamp post'' restriction of $m_{\tilde{t}_1} \leq 1.5$~TeV is removed, or 
rendered irrelevant by negative results of direct stop searches at some 
time in the future. We find that relaxation of this requirement does not 
affect the low-$m_0$ and low-$m_{1/2}$ region, but does populate the 
high-$m_0$ and high-$m_{1/2}$ ``decoupling'' region. At the same time, 
high values of $\tan\beta$ are marginally less constrained. Moreover, 
small negative values of $A_0$ are now allowed, but we should note that 
these correspond only to heavier values of $m_{\tilde{t}_1} > 1.5$~TeV. 
More interestingly, the scatter of points in both panels of 
Fig.~\ref{fig:decays} does not change appreciably. It follows that our 
remarks about the future of tests involving Higgs boson decays would 
generally remain valid even with a heavy $m_{\tilde{t}_1}$, which may be 
beyond the kinematic reach of the LHC.

To sum up, then, we have made a critical study of two of the most 
restrictive models of supersymmetry, cMSSM and NUHM, taking into account 
the recent measurements of the Higgs mass and the updates in many 
flavour-physics observables. We find that these models can actually 
survive all the constraints and predict a light stop which could be 
discovered, in principle, at the LHC. The allowed parameter ranges for 
both the models seem almost identical, except for $\tan\beta$, which is 
restricted to be $\lsim 45$ in the cMSSM with a light stop, while it 
could be larger in the NUHM. The branching ratios of the Higgs boson to 
$WW, ZZ, \tau\tau$ or $\gamma\gamma$ can be an important testing ground 
for the future of these models. However, as long as the experimental 
data do not exhibit definite disagreement with the SM, these SUSY models 
will continue to survive, and may even offer the exciting possibility of 
a light stop discovery in the coming runs of the LHC.

{\it Acknowledgements:} DG acknowledges support from ERC Ideas Starting 
Grant n. 279972 ``NPFlavour". He would also like to thank M.~Guchait 
for discussions.


\vfill

\end{document}